\documentstyle[11pt,amssymb,cite]{article}

\def\ben{\begin{equation}}
\def\een{\end{equation}}

  \let\n=\nu

\let\C=\Chi

\def\nn{\nonumber} \def\bd{\begin{document}} \def\ed{\end{document}}
\def\ds{\documentstyle} \let\fr=\frac \let\bl=\bigl \let\br=\bigr
\let\Br=\Bigr \let\Bl=\Bigl
\let\bm=\bibitem
\let\na=\nabla
\let\pa=\partial \let\ov=\overline
\newcommand{\be}{\begin{equation}}
\newcommand{\ee}{\end{equation}}
\def\ba{\begin{array}}
\def\ea{\end{array}}
\def\ft#1#2{{\textstyle{{\scriptstyle #1}\over {\scriptstyle #2}}}}
\def\fft#1#2{{#1 \over #2}}
\def\del{\partial}
\def\vp{\varphi}
\def\sst#1{{\scriptscriptstyle #1}}
\def\oneone{\rlap 1\mkern4mu{\rm l}}
\def\td{\tilde}
\def\wtd{\widetilde}
\def\ie{\rm i.e.\ }
\def\dalemb#1#2{{\vbox{\hrule height .#2pt
        \hbox{\vrule width.#2pt height#1pt \kern#1pt
                \vrule width.#2pt}
        \hrule height.#2pt}}}
\def\square{\mathord{\dalemb{6.8}{7}\hbox{\hskip1pt}}}
\newcommand{\ho}[1]{$\, ^{#1}$}
\newcommand{\hoch}[1]{$\, ^{#1}$}
\newcommand{\bea}{\begin{eqnarray}}
\newcommand{\eea}{\end{eqnarray}}
\newcommand{\ra}{\rightarrow}
\newcommand{\lra}{\longrightarrow}
\newcommand{\Lra}{\Leftrightarrow}
\newcommand{\ap}{\alpha^\prime}
\newcommand{\bp}{\tilde \beta^\prime}
\newcommand{\tr}{{\rm tr} }
\newcommand{\Tr}{{\rm Tr} }
\def\0{{\sst{(0)}}}
\def\1{{\sst{(1)}}}
\def\2{{\sst{(2)}}}
\def\3{{\sst{(3)}}}
\def\4{{\sst{(4)}}}
\def\5{{\sst{(5)}}}
\def\6{{\sst{(6)}}}
\def\7{{\sst{(7)}}}
\def\8{{\sst{(8)}}}
\def\n{{\sst{(n)}}}
\def\cA{{{\cal A}}}
\def\cB{{{\cal B}}}
\def\cF{{{\cal F}}}
\def\cH{{{\cal H}}}
\def\tV{\widetilde V}
\def\tW{\widetilde W}
\def\tH{\widetilde H}
\def\tE{\widetilde E}
\def\tF{\widetilde F}
\def\tA{\widetilde A}
\def\im{{i}}
\def\tY{{{\wtd Y}}}
\def\ep{{\epsilon}}
\def\vep{{\varepsilon}}
\def\R{\rlap{\rm I}\mkern3mu{\rm R}}
\def\bD{{{\bar D}}}

\def\R{\rlap{\rm I}\mkern3mu{\rm R}}
\def\bD{{{\bar D}}}
\def\R{{{\Bbb R}}}
\def\C{{{\Bbb C}}}
\def\H{{{\Bbb H}}}
\def\CP{{{\Bbb C}{\Bbb P}}}
\def\RP{{{\Bbb R}{\Bbb P}}}
\def\Z{{{\Bbb Z}}}
\def\bA{{{\Bbb A}}}
\def\bB{{{\Bbb B}}}
\def\bC{{{\Bbb C}}}
\def\bD{{{\Bbb D}}}
\def\bE{{{\Bbb E}}}
\def\bZ{{{\Bbb Z}}}
\def\Re{{{\frak{Re}}}}
\def\Im{{{\frak{Im}}}}
\def\cosec{{\,\hbox{cosec}\,}}
\def\Gm{{\Gamma_{\!\! -}}}
\def\Gp{{\Gamma_{\!\! +}}}
\def\stan{{standard }}
\def\nonstan{{supernumerary }}
\def\FF2{{ {}_{\sst 2}F_{\sst 1} }}
\def\FFF{{ {}_{\sst 3}F_{\sst 2} }}

\newcommand{\tamphys}{\it Center for Theoretical Physics,
Texas A\&M University, College Station, TX 77843}

\newcommand{\upenn}{\it Department of Physics and Astronomy,\\ University
of Pennsylvania, Philadelphia, PA 19104}

\newcommand{\brussels}{\it Physique Th\'eorique et Math\'ematique,
Universit\'e Libre de Bruxelles,\\ Campus Plaine C.P. 231, B-1050
Bruxelles, Belgium}

\newcommand{\auth}{ H. L\"u{\hoch 1}, Don N. Page{\hoch 2} and
C.N. Pope\hoch{1}}

\begin{document}
\begin{flushright}

Alberta Thy 06-04\ \ \ 
MIFP-04-04\\
{\bf hep-th/0403079}\\
March\  2004
\end{flushright}

\vspace{10pt}

\begin{center}

{\Large {\bf New Inhomogeneous Einstein Metrics on Sphere Bundles Over
            Einstein-K\"ahler Manifolds}}

\vspace{20pt}
\auth

\vspace{20pt}

{\it George P. and Cynthia W. Mitchell
Institute for Fundamental Physics,\\ Texas A\& M University,
College Station, TX 77843-4242, USA}


\vspace{40pt}

\underline{ABSTRACT}
\end{center}

    We construct new complete, compact, inhomogeneous Einstein metrics
on $S^{m+2}$ sphere bundles over $2n$-dimensional Einstein-K\"ahler
spaces $K_{2n}$, for all $n\ge1$ and all $m\ge 1$.  We also obtain
complete, compact, inhomogeneous Einstein metrics on warped products
of $S^m$ with $S^2$ bundles over $K_{2n}$, for $m>1$.  Additionally,
we construct new complete, non-compact Ricci-flat metrics with
topologies $S^m$ times $\R^2$ bundles over $K_{2n}$ that generalise
the higher-dimensional Taub-BOLT metrics, and with topologies $S^m
\times \R^{2n+2}$ that generalise the higher-dimensional Taub-NUT
metrics, again for $m>1$.

{\vfill\leftline{}\vfill \vskip 10pt \footnoterule 
{\footnotesize {\hoch 1}
Research supported in part by DOE grant
DE-FG03-95ER40917.}

{\footnotesize
{\hoch 2}
Permanent address: Theoretical Physics Institute, 412 Physics Lab., University
of Alberta, Edmonton,}

{\footnotesize $\phantom *$ Alberta T6G 2J1, Canada.  Research
  supported in part by the Natural Science and Engineering }

{\footnotesize $\phantom *$ Research
  Council of Canada.}
}

\pagebreak
\setcounter{page}{1}

\newpage

\section{Introduction}

   Compact homogeneous Einstein spaces, in the form of cosets $G/H$,
are well known.  Less well known are the examples of complete, compact
Einstein spaces that are inhomogeneous.  The first explicit example
was obtained in \cite{page}, by considering a limit of the
four-dimensional Euclidean Kerr-de Sitter solution.  It has the
topology of the non-trivial $S^2$ bundle over $S^2$, which is $\CP^2
\# \overline{\CP^2}$.  A wide class of complete, compact inhomogeneous
Einstein metrics in arbitrary even dimensions $d=2n+2$ was then
obtained \cite{berber,pagpop1}.  These are defined on manifolds which
are $S^2$ bundles over $K_{2n}$, where $K_{2n}$ is an
Einstein-K\"ahler space.  The original example in \cite{page} is the
special case where $n=1$, with $K_2$ then being the 2-sphere with its
standard Einstein-K\"ahler metric.
 
    Further explicit examples of complete, compact, inhomogeneous
Einstein spaces have recently been obtained \cite{hassakyas}.  Like the
original example in \cite{page}, these were again found by considering
limits of Kerr-de Sitter black holes, this time in $d\ge 5$
dimensions.  Specifically, in $d=5$ the general Kerr-de Sitter black
hole with two independent rotation parameters is known
\cite{hawhuntay}.  By taking limiting cases of the
Euclideanised black-hole solutions, it was shown in \cite{hassakyas}
that these metrics encompass a countable infinity of inequivalent
complete, compact Einstein metrics.  The topologies of the associated 
smooth manifolds are either $S^3\times S^2$ or the non-trivial $S^3$
bundle over $S^2$, according to whether a certain integer is even or odd
\cite{hassakyas}.  In addition, further complete, compact inhomogeneous
metrics were constructed in all dimensions $d>5$ in \cite{hassakyas}.  
These metrics, one for each dimension $d>5$, were obtained by taking 
a limit of the Euclideanised $d$-dimensional 
Kerr-de Sitter black hole with one rotation
parameter.\footnote{Examples of new, inhomogeneous Einstein-Sasaki
metrics in five \cite{gaunt1} and higher \cite{gaunt2} dimensions 
have been obtained recently too.}  The topology
of the manifold is the non-trivial $S^{d-2}$ bundle over $S^2$
\cite{hassakyas}.

   In this paper, we shall give a construction of complete, compact,
inhomogeneous Einstein metrics on $S^{m+2}$ bundles over Einstein-K\"ahler
spaces $K_{2n}$ of any (even) dimension $2n\ge 2$.  The metrics on the 
$S^{d-2}$ bundles over $S^2$ obtained in \cite{hassakyas} are thus
special cases of our new metrics, when the Einstein-K\"ahler base is taken
to be $S^2$.  Our construction is also a natural generalisation of the one
in \cite{berber,pagpop1}.

    We also obtain a new infinite class of complete, compact,
inhomogeneous Einstein metrics on warped products of $S^m$ with $S^2$
bundles over $K_{2n}$, for $m > 1$.   These are different
from the direct-product Einstein metrics that one could trivially
construct from the previous results in \cite{berber,pagpop1}.

   In addition to the new compact Einstein metrics described above, we
also obtain new complete, non-compact Ricci-flat metrics that
generalise the Taub-NUT and Taub-BOLT metrics in the literature.  
These have the topologies $S^m \times \R^{n+2}$, and $S^m$ times $\R^2$ 
bundles over $K_{2n}$ respectively, with $m>1$.

\section{The Local Solutions}\label{localsolsec}

   Let $d\Sigma_{2n}^2$ be an Einstein-K\"ahler metric of real dimension
$2n$, normalised such that
\be
R_{ab} = \lambda\, g_{ab}\,.
\ee
In this paper, we shall be concentrating on the case where $\lambda>0$. 
We then write the following ansatz for metrics of dimension $d=2n+m+2$:
\be
d\hat s^2 = dt^2 + \beta^2\, (d\tau- 2A)^2 + \gamma^2\, d\Sigma_{2n}^2 
+ \delta^2 \, d\Omega_m^2\,,\label{einsttot0}
\ee
where $\beta$, $\gamma$ and $\delta$ are functions of $t$,
$d\Omega_m^2$ is the standard metric on the unit sphere $S^m$, and $A$
is a potential for the K\"ahler form on $d\Sigma_{2n}^2$: $J=dA$.  This 
metric is similar to the one considered in \cite{pagpop1}, except that
now an extra $m$ dimensions involving the $m$-sphere are included.  A
straightforward calculation shows that in the orthonormal frame
$\hat e^0 =dt$, $\hat e^1 = \beta\,  (d\tau- 2A)$, $\hat e^a = \gamma\, 
e^a$, $\hat e^i = \delta\, e^i$, the non-vanishing components of the
Ricci tensor are given by
\bea
\hat R_{00} &=& - \fft{\ddot \beta}{\beta} - 2n\, 
   \fft{\ddot\gamma}{\gamma} - m\, \fft{\ddot \delta}{\delta}\,,\nn\\
\hat R_{11} &=& - \fft{\ddot \beta}{\beta} - 2n\, \fft{\dot \beta \, 
\dot \gamma}{\beta\, \gamma} - m\, 
\fft{\dot \beta\, \dot\delta}{\beta\, \delta} + 
   2n\, \fft{\beta^2}{\gamma^4}\,,\nn\\
\hat R_{ab} &=& -\Big(\fft{\ddot \gamma}{\gamma} + 
   \fft{\dot\beta\,\dot\gamma}{\beta\,\gamma} + m\, \fft{\dot\gamma\, 
         \dot\delta}{\gamma\, \delta} + (2n-1)\, 
            \fft{{\dot\gamma}^2}{\gamma^2} + \fft{2 \beta^2}{\gamma^4}
              -\fft{\lambda}{\gamma^2} \Big)\, \delta_{ab}\,,\nn\\
\hat R_{ij} &=& -\Big( \fft{\ddot\delta}{\delta} + 
                \fft{\dot\beta\, \dot \delta}{\beta\, \delta} +
                 2n\, \fft{\dot\gamma\, \dot\delta}{\gamma\, \delta} 
                 + (m-1)\, \fft{{\dot\delta}^2}{\delta^2} - 
                      \fft{m-1}{\delta^2}\Big)\, \delta_{ij}\,.
\eea

  One can look for solutions to the Einstein equation
\be
\hat R_{AB} = \Lambda\, \delta_{AB}\label{einsttoteq}
\ee
by following a procedure similar to the one described in
\cite{pagpop1}.  We find that a solution is given by
\be
d\hat s^2 = \fft{(1-r^2)^n\, dr^2}{P(r)} + \fft{c^2\, P(r)}{(1-r^2)^n}\, 
(d\tau -2A)^2 + c\, (1-r^2)\, d\Sigma_{2n}^2 + 
\fft{(m-1)}{\Lambda-\lambda\, c^{-1}}\, r^2\, d\Omega_m^2\,,
\label{einsttot}
\ee
where $c$ is an arbitrary constant of integration and the function $P(r)$
is given by
\be
\fft{d(r^{m-1}\, P(r))}{dr}  = r^{m-2}\, [ \Lambda\, (1-r^2)^{n+1} 
              - \lambda\, c^{-1}\, (1-r^2)^n]\,.\label{Pder}
\ee
Thus we have
\be
P(r) = \fft{\Lambda}{m-1}\, \FF2(-n-1, \ft{m-1}{2}; \ft{m+1}{2}; r^2)-
              \fft{\lambda\, c^{-1}}{m-1}\, 
                  \FF2(-n, \ft{m-1}{2}; \ft{m+1}{2}; r^2) + \mu\,
                r^{1-m}\,,\label{psolm}
\ee
where $\FF2$ denotes the standard hypergeometric function.  
The quantity $\mu$ in the final term is a further constant of
integration.  The rest of the terms in $P(r)$ form a polynomial in $r$
of degree $2n$.  It is useful to note that the hypergeometric functions
that arise here have the form
\bea
\fft1{m-1}\, \FF2(-1,\ft{m-1}{2}; \ft{m+1}{2}; r^2) &=& \fft1{m-1} 
   - \fft{r^2}{m+1}\,,\nn\\
\fft1{m-1}\, \FF2(-2,\ft{m-1}{2}; \ft{m+1}{2}; r^2) &=& \fft1{m-1} 
   - \fft{2r^2}{m+1} + \fft{r^4}{m+3}\,,\nn\\
\fft1{m-1}\, \FF2(-3,\ft{m-1}{2}; \ft{m+1}{2}; r^2) &=& \fft1{m-1} 
   - \fft{3r^2}{m+1} +\fft{3r^4}{m+3} - \fft{r^6}{m+5}\,,
\eea
{\it etc}., with the pattern continuing to higher values of $n$ in the
obvious way.

   There are in total two non-trivial parameters in the Einstein
metrics (\ref{einsttot}). One is the integration constant $\mu$, and
the other can be taken to be the dimensionless constant
\be
\nu\equiv \fft{c\, \Lambda}{\lambda}\,.\label{nudef}
\ee
The constant $\Lambda$ is just the overall
scale-setting Einstein constant in (\ref{einsttoteq}), and it can
easily be seen by rescaling the Einstein-K\"ahler base metric that
$\lambda$ and $c$ always occur in the combination $\lambda\, c^{-1}$.
We shall make various convenient choices for the ``trivial''
parameters $\Lambda$ and $\lambda$ as the occasion arises.

    The case $m=1$ is degenerate in the above
parameterisation. However, it can easily be obtained as a limiting
case, in which one absorbs the $(m-1)$ factor into a rescaling of
$d\Omega_m^2$ before setting $m=1$, and so we have, with 
$0\le \chi\le 2\pi$ forming an $S^1$,
\be
d\hat s^2 = \fft{(1-r^2)^n\, dr^2}{P(r)} + \fft{c^2\, P(r)}{(1-r^2)^n}\, 
(d\tau -2A)^2 + c\, (1-r^2)\, d\Sigma_{2n}^2 + 
   P_0^{-1}\,  r^2\, d\chi^2\,,
\label{einsttotm1}
\ee
where $P_0$ is an arbitrary positive constant.  The Einstein equations
now require $\nu=1$ and that $P(r)$ satisfies $P'(r) = -\Lambda\, r\, 
(1-r^2)^n$ and hence $P(r)$ is given by
\be
P(r) = \fft{\Lambda}{2 (n+1)}\, 
   [(1-r^2)^{n+1}-1] + \mu\,.\label{logfree}
\ee
where $\mu$ is an integration constant.  
The metric for $m=1$ thus has two non-trivial parameters, namely 
$\mu$ and $P_0$.

   In certain low-dimensional cases, namely $n=1$ with $m=1$, 2 and 3,
and $n=2$ with $m=1$, some local solutions that overlap 
with ours were obtained in \cite{manste}.

   The local Einstein metrics we have obtained are in general
singular, in the sense that they cannot extend onto smooth manifolds.
For special choices of the parameters, however, non-singular metrics
arise. The issue of whether this occurs or not can be understood by
studying the local forms of the metrics at the ``endpoints'' of the
range of the radial coordinate $r$.  The idea is that $r$ ranges
between two adjacent values, say $r_1$ and $r_2$, at which one or more
of the metric functions $\beta$, $\gamma$ or $\delta$ in
(\ref{einsttot0}) vanishes.  At such points, one can obtain regular
behaviour if it should happen that the vanishing metric functions
imply the collapse of spheres to an origin of spherical polar
coordinates.  If the collapsing sphere has dimension greater than 1,
the Einstein equations imply that the rate of collapse will
automatically be that needed for regularity.  In the case of a
one-dimensional sphere, \ie a circle, $S^1$, its rate of collapse is
not governed by the Einstein equations, and it places a condition on
the periodicity of the $S^1$ coordinate in order to avoid a conical
singularity.

\section{Complete Metrics on $S^{m+2}$ Bundles Over $K_{2n}$}\label{sm2bun}

   In this section, we study the circumstances under which the above
local solutions extend smoothly onto manifolds that are $S^{m+2}$
bundles over the $K_{2n}$ Einstein-K\"ahler base spaces.  In this case
one endpoint for the radial coordinate is $r=r_1=0$, signalling a
collapse of the sphere $S^m$, whose metric is $d\Omega_m^2$ in
(\ref{einsttot}) or $d\chi^2$ in (\ref{einsttotm1}).  Without loss
of generality we can take $r\ge0$, so that the upper endpoint will
occur at $r=r_2$, the first positive root of the function $P(r)$,
where $\beta=0$:
\be
P(r_2)=0\,.
\ee
The cases $m=1$ and $m>1$ require separate discussions, and we shall 
consider the latter first.

\subsection{$m>1$: $S^{m+2}$ bundles over $K_{2n}$}\label{mg1sec}

   In order to have regular solutions, the function $P(r)$ must be
regular at $r=0$, which for $m>1$ implies that the integration
constant $\mu$ in (\ref{psolm}) must vanish, and hence $P(r)$ is a
polynomial in $r$ of degree $2n$.  As discussed above, the Einstein
equations automatically imply in this case that the spheres $S^m$
collapse in a regular fashion at $r=0$.  The question of regularity
thus devolves upon the behaviour at the first positive zero $r_2$ of
the polynomial $P(r)$, at which the length of the $U(1)$ fibres
parameterised by $\tau$ goes to zero.

   Before considering the regularity conditions at $r=r_2$, it should
noted that the possible periods for the $U(1)$ fibre coordinate $\tau$
are dictated by the first Chern class $c_1$ of the tangent bundle of
the Einstein-K\"ahler space $K_{2n}$.  This was discussed in detail in
\cite{pagpop1}, and here we shall just summarise the conclusion.  The
allowed periods for $\tau$ are given by
\be
\Delta\tau = \fft{4\pi\, p}{k\, \lambda}\,,\label{tauperiods}
\ee
where $k$ is any positive integer, and $p$ is a non-negative integer,
defined as that integer such that $c_1$ evaluated on $H_2(K_{2n}, \Z)$
is $\Z \cdot p$, \ie the integers divisible by $p$.  Amongst all the 
Einstein-K\"ahler manifolds $K_{2n}$, the integer $p$ is maximised in 
$\CP^n$, for which $p=n+1$.   

  In order for the metric (\ref{einsttot}) to be regular at $r=r_2$,
it must be the case that the required periodicity for $\tau$ is
contained within the set of allowed values given in
(\ref{tauperiods}).  By writing $P(r)\sim -P'(r_2)\, (r_2-r) =
- P'(r_2)\, \rho^2$ near $r=r_2$, one finds that the metric
(\ref{einsttot}) restricted to the $(\rho,\tau)$ plane has no conical
singularity if $\tau$ has period given by
\be
\Delta\tau = \fft{4\pi\, (1-r_2^2)^n}{c\, (-P'(r_2))}\,,
\ee
where $P'(r)$ is the derivative of $P(r)$ with respect to $r$, which is 
negative at $r=r_2$.  Combining 
this with (\ref{tauperiods}), and making use of (\ref{Pder}), we therefore
obtain the condition
\be
\fft{k}{p} = \fft{1-\nu\, (1-r_2^2)}{r_2}\,.\label{r2con}
\ee
Note that now $\nu\equiv c\, \Lambda/\lambda$ is the single
remaining non-trivial parameter in the Einstein metrics
(\ref{einsttot}), since we have already set $\mu=0$ for
regularity at $r=0$.  The solution (\ref{psolm}) for $P(r)$, with
$\mu=0$, implies that $P(r_2)=0$ gives
\be
\nu = \fft{\FF2(-n,\ft{m-1}{2}; \ft{m+1}{2}; r_2^2)}{
          \FF2(-n-1,\ft{m-1}{2}; \ft{m+1}{2}; r_2^2)}\,,\label{nueq}
\ee
and the regularity condition (\ref{r2con}) then gives
\be
\fft{k}{p} = \kappa(r_2) \equiv \fft1{r_2}\, \Big[
                  1 - (1-r_2)^2\, 
                \fft{\FF2(-n,\ft{m-1}{2}; \ft{m+1}{2}; r_2^2)}{
          \FF2(-n-1,\ft{m-1}{2}; \ft{m+1}{2}; r_2^2)}\Big]\,,\label{kappa}
\ee
which is the one non-trivial condition (determining $r_2$) for a
regular metric.

   For small $r_2$ we find that $\kappa(r_2)\sim 2r_2/(m+1)$, and it
is obvious from (\ref{kappa}) that $\kappa(1)=1$.  It follows from the 
continuity of $\kappa(r)$ that for any integer $k$ in the range
\be
 0 <k < p \,,\label{krange}
\ee
we can find a corresponding $r_2$ with $0<r_2<1$ that satisfies
(\ref{kappa}).  Equation (\ref{nueq}) then determines the value of the
non-trivial parameter $\nu$.  Thus we have complete, nonsingular Einstein
metrics for all integer $k$ within the range (\ref{krange}).  

    Over each point in the Einstein-K\"ahler base space $K_{2n}$, the
metric (\ref{einsttot}) describes a sphere $S^{m+2}$, foliated by
$S^1\times S^m$ surfaces at each value of $r$ with $0< r < r_2$.  The
$S^m$ degenerates at $r=r_1=0$, and the $S^1\sim U(1)$ degenerates at
$r=r_2$. Since the $U(1)$ is fibred over the $K_{2n}$ base, it follows
that the total space is an $S^{m+2}$ bundle over $K_{2n}$.

   The special cases where $n=1$, for which the Einstein-K\"ahler base
space $K_2$ is just $S^2$, reproduce the Einstein metrics on $S^{d-2}$ 
bundles over $S^2$ that were obtained in \cite{hassakyas}.  Since $p=2$
in this case, it follows from (\ref{krange}) that the integer $k$ can
only take the value $k=1$.  In this case, the function $P$ is given by
\be
P=\Lambda\, \Bigl[\fft{1}{m-1} - \fft{2r^2}{m+1} + \fft{r^4}{m+3} -
\fft{1-r_2^2}{1-k\,r_2/p}\, \Bigl( \fft{1}{m-1} -
\fft{r^2}{m+1}\Bigr)\Bigr]\,,
\ee
with the constant $r_2$ satisfying the regularity condition
\be
\fft{1}{m-1} - \fft{2r_2^2}{m+1} + \fft{r_2^4}{m+3} =
4r_2\, \Bigl[\fft{1}{(m-1)(m+1)} - \fft{r_2^2}{(m+1)(m+3)}
\Bigr]\,.
\ee
For $n=2$, we have that $p=3$, and hence $k=1$ or $k=2$.  The function
$P$ is given by
\be
P=\Lambda\, \Bigl[ \fft{1}{m-1} - \fft{3r^2}{m+1} +
\fft{3r^4}{m+3} - \fft{r^6}{m+5} -
\fft{1-r_2^2}{1-k\, r_2/p}\, \Bigl(
\fft{1}{m-1} - \fft{2r_2^2}{m+1} + \fft{r^4}{m+3}\Bigr)\Bigr]\,,
\ee
where the constant $r_2$ satisfies the regularity condition
\bea
&&\fft{1}{m-1} - \fft{3r_2^2}{m+1} + \fft{3r_2^4}{m+3} -
\fft{r_2^6}{m+5} =\nn\\
&&\qquad\qquad\qquad \fft{2p}{k}\Bigl[
\fft{1}{(m-1)(m+1)} - \fft{2r_2^2}{(m+1)(m+3)} +
\fft{r_2^4}{(m+3)(m+5)}\Bigr]\,.
\eea
As an explicit example, let us consider $m=2$.  We have
$r_2\approx  0.445844 $ when $k=1$, and
$r_2\approx 0.750285$ when $k=2$.

\subsection{$m=1$: $S^3$ bundles over $K_{2n}$}   

   We now consider the case where $m=1$, which was not covered in 
section \ref{mg1sec}.   The function $P(r)$ is then given by
(\ref{logfree}).  The regularity condition (\ref{r2con}) then implies
that
\be
\fft{k}{p} = r_2\,.
\ee
Thus we can have any $k$ in the range $0< k < p$, as in (\ref{krange}),
with the root $r_2$ lying in the interval $0<r_2 <1$.
Setting $P(r_2)=0$ in (\ref{logfree}) determines $\mu$, and hence gives
\be
P(r) = (1-r^2)^{n+1} - (1-k^2/p^2)^{n+1}\,,\label{psolmm1}
\ee
where for simplicity, we have now made the conventional scale choice
$\Lambda=2n+2$.  With our choice that the coordinate $\chi$ on $S^1$
in the metric (\ref{einsttotm1}) has period $2\pi$, the regularity
condition for no conical singularity at $r=0$ implies that the
constant $P_0$ in (\ref{einsttotm1}) is given by $P_0=P(0)$.

   In summary, we have complete, compact, inhomogeneous Einstein
metrics for all $k$ in the range (\ref{krange}), on spaces whose
topologies are $S^3$ bundles over the Einstein-K\"ahler base space
$K_{2n}$.  Interestingly, these metrics are all fully explicit, 
with 
\be
d\hat s_{2n+3}^2 = \fft{(1-r^2)^n\, dr^2}{P(r)} + 
\fft{P(r)}{(1-r^2)^n}\, (d\tau-2A)^2 + (1-r^2)\, d\Sigma_{2n}^2 
    + P_0^{-1}\, r^2\, d\chi^2\,,
\ee
where $P(r)$ is given by (\ref{psolmm1}), and here we have normalised
the metric $d\Sigma_{2n}^2$ on the Einstein-K\"ahler base space
$K_{2n}$ so that $\lambda =2n+2$.  The special case with $n=1$, 
for which $K_2=S^2$, was found in \cite{hassakyas}.

\section{Warped-Product Einstein Metrics} 

    In this section we consider a different class of complete Einstein
metrics encompassed within the solutions (\ref{einsttot}) and
(\ref{einsttotm1}), in which the radial coordinate $r$ ranges between
two adjacent positive zeroes of $P(r)$, at $r_1$ and $r_2$.  In order
also to avoid the singularity of the metric components at $r=1$, we 
require
\be
0 < r_1 < r_2 <1\,.\label{r1r2range}
\ee
It is evident from the solution for $P(r)$ given in (\ref{logfree}), 
which is monotonically decreasing as $r$ increases from 0 to 1, that
we cannot achieve two roots as in (\ref{r1r2range}) when $m=1$.  Thus
to obtain warped-product Einstein metrics we must restrict attention
to the cases with $m>1$.

    Since $P(r)$ now vanishes at both endpoints, $r_1$ and $r_2$, of
the range of the radial coordinate, there is a regularity condition of
the form (\ref{r2con}) at each endpoint.  Specifically, we now have
\be
\fft{\nu\, (1-r_1^2) -1}{r_1}= \fft{k}{p} = \fft{1-\nu\, (1-r_2^2)}{r_2}\,,
\label{r2con2}
\ee
where the reversal of sign in the condition at $r_1$ results from the 
fact that $P'(r_1)$ is positive, whereas $P'(r_2)$ is negative.  Thus
we have
\bea
\nu &=& \fft1{1 - r_1\, r_2}\,,\\
\fft{k}{p} &=& \fft{r_2-r_1}{1-r_1\, r_2}\,.\label{kp2}
\eea
It follows immediately from (\ref{kp2}) that $k/p>0$ and $1-k/p >0$, and
hence $k$ must lie in the range $0< k < p$, as in (\ref{krange}).

   By definition, the endpoints $r_1$ and $r_2$ satisfy $P(r_i)=0$. We
can easily arrange that $P(r_i)=0$ for one endpoint by choosing the
non-trivial parameter $\mu$ appropriately in (\ref{psolm}).  For the
$P(r_i)=0$ at the other endpoint, we see that the integral of the
right-hand side of (\ref{Pder}) between the endpoints must vanish,
which implies
\be
D(r_1,r_2)\equiv 
\int_{r_1}^{r_2} dr\, r^{m-2}\, (1-r^2)^n\, (r_1\, r_2 -r^2) =0\,.
\label{r1r2int}
\ee

   For a given $k/p$, (\ref{kp2}) and (\ref{r1r2int}) give two
algebraic equations for the unknowns $r_1$ and $r_2$.  We shall show
that for each $k$ in the range $0<k<p$, there always exists a solution
for $r_1$ and $r_2$ satisfying (\ref{r1r2range}), and hence we have a
complete, nonsingular Einstein space for each such $k$.

   To show the existence of solutions to (\ref{kp2}) and
(\ref{r1r2int}), it is helpful for now to view $R\equiv k/p$ as a 
continuous variable that satisfies
\be
0 \le R\le 1\,.
\ee
Of course it is actually discrete, and the limits $R=0$ and $R=1$ are 
disallowed by (\ref{krange}), but it is easier first to discuss the 
existence of solutions to 
\be
R= \fft{r_2-r_1}{1-r_1\, r_2}\,,\qquad D(r_1,r_2)=0\,,\label{RD}
\ee
for $0\le r_1\le r_2 \le 1$.  

    We first consider the function
$D(r_1,r_2)$ with $r_2=1$, which, from (\ref{RD}), implies $R=1$.
Clearly, $D(r_1,1)$ is negative if $r_1=0$, whilst for $r_1$ 
close to 1 an expansion for $r_1=1-\epsilon$ shows that
$D(r_1,1)$ is positive.  By continuity, there therefore exists some
$r_1$ for which $D(r_1,1)=0$.  If $r_2$ is reduced from $r_2=1$, then $R$ 
reduces too.  $D(r_1,r_2)$ is again negative if $r_1=0$, whilst it is
positive if $r_1$ is infinitesimally below $r_2$, provided that
\be
r_2^2 > \fft{m}{m+2n}\,.\label{r2bound}
\ee
By continuity, there is therefore always a solution
of $D(r_1,r_2)=0$ provided that (\ref{r2bound}) is satisfied.  In fact, if
$r_2^2$ approaches $m/(m+2n)$, then the solution to $D(r_1,r_2)=0$ gives
$r_1$ approaching $r_2$, and hence, from (\ref{RD}), $R$ approaches 0.
Thus for any value of $R$ with $0<R<1$ we have a solution for $r_1$
and $r_2$ with $0< r_1 < r_2 <1$.  

   We have therefore established that for all positive integers $m$,
we have a complete, compact, nonsingular Einstein metric for each
integer $k$ with $0<k<p$.
   
   Since the coefficient of $d\Omega_m^2$ in (\ref{einsttot}) is
non-vanishing for the entire range of the radial coordinate, $r_1\le r
\le r_2$, it follows that topologically the manifolds for these
metrics are simply the product of $S^m$ with the same class of $S^2$
bundles over $K_{2n}$ that was discussed in \cite{pagpop1}.
Metrically, however, the Einstein spaces we have obtained here are
{\it warped} products of $S^m$ with the $S^2$ bundles over $K_{2n}$,
since the size of the $S^m$ varies with $r$.

   Since the size of $S^m$ is non-zero over the complete coordinate
range $r_1 \le r\le r_2$, we do not need to restrict $d\Omega_m^2$ to
be the Einstein metric on the unit $S^m$.  We can take any Einstein
space with $R_{ij} = (m-1)\, \delta_{ij}$ in place of $S^m$. This, of
course, is different from the situation in section \ref{sm2bun}, where
the coefficient of $d\Omega_m^2$ vanished at the lower endpoint, and
so it necessarily had to be a round $m$-sphere for regularity at
$r=0$.

    It should be emphasised that the Einstein spaces we have
obtained in this section are distinct from examples with identical
topology that we could trivially obtain by taking the direct product
of the $S^2$ bundles over $K_{2n}$ in \cite{pagpop1} with
appropriately-scaled metrics on $S^m$, or on any other Einstein space
with positive Ricci tensor in place of $S^m$.  Thus for every
direct-product metric of this type, our new construction in this
section yields an inequivalent warped-product Einstein space with the
same topology.  

   Now let us present some explicit examples. First,
 consider the cases with $n=1$ and $m\ge 2$.  The function $P$
is given by
\bea
P&=&\Lambda\,\Big[\fft{r^4}{m+3} -
\fft{(r_1r_2+1)\, r^2}{m+1} + \fft{r_1r_2}{m-1}\nn\\
&&\qquad\qquad\qquad -
(\fft{r_1}{r})^{m-1}\, (\fft{r_1^4}{m+3} -
\fft{(r_1r_2+1)\, r_1^2}{m+1} + \fft{r_1r_2}{m-1})\Bigr]\,.
\eea
For a specific example consider the six-dimensional metric obtained by
taking $m=2$.  The regularity conditions (\ref{r1r2int}) and (\ref{kp2})
with $k=1$, $p=1$ imply that
\bea
r_1 &=&\fft{\sqrt{4\sqrt{85} -19 } +\sqrt{85} -10  }6
\approx 0.574634\,,\nn\\
r_2 &=&\fft{\sqrt{4\sqrt{85}-19 } -\sqrt{85}  +10 }6 \approx
0.834786\,.
\eea
Thus we have $\nu=1 + \sqrt{17/20}\approx 1.92195$.
For the seven-dimensional example with $m=3$, we have
\bea
r_1 &=& \fft{\sqrt{6\sqrt{21} -18 }+\sqrt{21}- 5  }{4}\approx
 0.666011\,,\nn\\
r_2 &=& \fft{\sqrt{6\sqrt{21} -18 }-\sqrt{21} + 5 }{4}\approx
0.874723\,,
\eea
and $\nu=\ft14(5 + \sqrt{21})\approx 2.39564$.

\section{Non-Compact Ricci-Flat Metrics}

    Here, we show that one can obtain complete, nonsingular,
non-compact Ricci-flat metrics from the solutions.  Setting
$\Lambda=0$, $c=-1$ and $P(r) = (-1)^n\, \wtd P(r)$ in
(\ref{einsttot0}) and (\ref{einsttotm1}), we find that when $m=1$
there are no non-singular solutions within the ansatz that we are
considering, and thus we shall concentrate on the cases with $m>1$.
For these, the metric (\ref{einsttot}) becomes
\be
d\hat s^2 = \fft{(r^2-1)^n\, dr^2 }{\wtd P(r)} + 
  \fft{\wtd P(r)}{(r^2-1)^n}\, (d\tau -2A)^2 + (r^2-1)\,
  d\Sigma_{2n}^2 + \fft{m-1}{\lambda}\, r^2\, d\Omega_m^2\,,
\label{taubbolt}
\ee
which is Ricci flat, where
\bea
\wtd P(r) &=& \lambda \, r^{1-m}\, \int_{r_1}^r dx\, x^{m-2}\,
(x^2-1)^n \label{ptilde}\\
&=& \fft{(-1)^n\, \lambda}{m-1}\, \Big[\FF2(-n,
  \ft{m-1}2;\ft{m+1}2;r^2) - \Big(\fft{r_1}{r}\Big)^{m-1}\, 
      \FF2(-n,\ft{m-1}2;\ft{m+1}2;r_1^2)\Big]\,.
\eea

     The radial coordinate runs for $r=r_1$ to
$r=\infty$.  There are two cases to consider, depending upon whether
$r_1=1$ or $r_1>1$.

\subsection{$r_1>1$: Generalised Taub-BOLT metrics}\label{taubboltsec}

   Let us first consider the case $r_1>1$.  Since $\Lambda=0$ we
have from (\ref{nudef}) that $\nu=0$ and hence, analogously to
(\ref{r2con}), regularity at $r_1=1$ implies
\be
\fft{k}{p}= \fft1{r_1}\,.
\ee
Again, therefore, we have that regularity requires $0<k<p$, as in
(\ref{krange}).  Thus for each allowed value of $k$ we have a
complete, nonsingular, non-compact Ricci-flat metric with $r_1=p/k$.  
Topologically, the manifold is $S^m$ times an $\R^2$ bundle over the
Einstein-K\"ahler base space $K_{2n}$.  At large distance, we have
\be
\wtd P(r) \sim \fft{\lambda}{2n+m-1}\, r^{2n}\,,
\ee
and so the coefficient of $(d\tau-2A)^2$ in (\ref{taubbolt}) becomes
asymptotically constant.  This is the same type of asymptotic
behaviour that is encountered in the Taub-NUT and Taub-BOLT metrics.
In fact, the metrics with $r_1>1$ that we have obtained here are 
generalisations of the standard Taub-BOLT metrics obtained in
\cite{page2,baisbatt}.  The new feature is the added $S^m$ factor. 

    It should be remarked that since the coefficient of $d\Omega_m^2$
remains non-zero in the entire range of the radial variable, we can
replace the sphere $S^m$ by any other Einstein space, normalised such
that $R_{ij} = (m-1)\, \delta_{ij}$.

    Some explicit low-dimensional examples are as follows.  We set
$\lambda=1$ for convenience.  For $n=1$, with $m>1$, we have
\bea
d\hat s^2_{m+4} &=&\fft{(m+1)\, dr^2}{U} +
 \fft{4U}{m+1}\, \,(d\psi+\cos\theta\, d\phi)^2 + 
(r^2-1)\, (d\theta^2+\sin^2\theta\, d\phi^2) \nn\\
&&+ (m-1)\, r^2\,
 d\Omega_m^2\,,\nn\\
U &=& 1-\fft{2+ 2^{m-1}\, (3m-5)\,    r^{1-m}}{(m-1)\,
    (r^2-1)}\,. 
\eea
The case $n=1$, $m=2$ is of some interest, giving
\be
d\hat s_6^2 = \fft{3 r\, (r-1)\, dr^2}{(r+1)\, (r-2)}
     + \fft{4(r+1)\, (r-2)}{3 r\, (r-1)}\, (d\psi+\cos\theta\, d\phi)^2
+ (r^2-1)\, (d\theta^2+\sin^2\theta\, d\phi^2) + r^2\, d\Omega_2^2\,.
\label{tbolt}
\ee
We shall return to this in section (\ref{taubnutsec}).

   For $n=2$, with $m>1$, we have
\bea
d\hat s_{m+6}^2 &=& \fft{(m+3)\, dr^2}{U} + \fft{U}{m+3}\, 
(d\tau-2A)^2 + (r^2-1)\,
d\Sigma_4^2 + (m-1)\, r^2\, d\Omega_m\,,\nn\\
U &=& 1 - \fft{1}{(m^2-1)(r^2-1)^2}\,\Bigl[
4(m-1)r^2 -4(m+1)\\
&&\qquad\qquad+ [(m^2-1) r_1^4 -2(m-1)(m+3) r_1^2
+ (m+1)(m+3)](r_1/r)^{m-1}\Bigr]\,,\nn
\eea
where $r_1$ is 3 or $\ft32$, corresponding to $k=1$ or $k=2$
respectively, and $d\Sigma_4^2$ is the Fubini-Study metric on $\CP^2$, 
scaled so that $\lambda=1$.

\subsection{$r_1=1$: Generalised Taub-NUT metrics}\label{taubnutsec}
 
    Next, we consider the case when $r_1=1$.  The discussion of
regularity at $r=1$ is different from the previous one, since now the
coefficient of $d\Sigma_{2n}^2$ is also going to zero.  Regularity is
achieved if and only if $K_{2n}$ is taken to be $\CP^n$ with
$d\Sigma_{2n}^2$ its standard Fubini-Study metric.  The integer $k$ in
(\ref{tauperiods}) must be $k=1$.  Setting $r_1=1$
in (\ref{ptilde}), we find that near $r=1$, $\wtd P(r)$ goes as
\be
\wtd P(r) \sim \fft{2^n\, \lambda }{n+1}\, (r-1)^{n+1}\,.
\ee
Setting $r-1=\rho^2$, and taking $\lambda=2n+2$ for convenience (this 
gives the canonical ``unit'' size for $\CP^n$), we find that near
$\rho=0$ the metric (\ref{taubbolt}) for $m>1$ takes the regular form
\be
d\hat s^2 \sim 2(d\rho^2 + \rho^2\, d\wtd \Omega_{2n+1}^2) + 
\fft{m-1}{2n+2}\, d\Omega_m^2\,.
\ee
Here, $d\Omega_{2n+1}^2 = (d\tau-2A)^2 + d\Sigma_{2n}^2$ is the metric
on the unit $S^{2n+1}$, arising as the Hopf fibration over $\CP^n$.
At large distance, on the other hand, the asymptotic form of the metrics
(\ref{taubbolt}) with $r_1=1$ is the same as in our previous discussion
in section \ref{taubboltsec}.  Topologically, these new metrics are
non-singular on the manifolds $S^m\times \R^{2n+2}$.

   The metrics that we obtain in this case with $r_1=1$ are
generalisations of the Taub-NUT metrics obtained in higher dimensions
in \cite{baisbatt}.  The new feature is again the addition of the
$S^m$ factor.  As in the generalised Taub-BOLT metrics in section
\ref{taubboltsec}, since the coefficient of $d\Omega_m^2$ never
vanishes for $r\ge 1$, we can replace the sphere $S^m$ by any Einstein
space with $R_{ij} = (m-1)\, \delta_{ij}$.  

    Finally, we remark that our discussion of Ricci-flat metrics in this 
section can be straightforwardly extended to give complete, non-singular,
non-compact metrics with negative cosmological constant $\Lambda$.  Since
all the formulae can be immediately read off from the results in section
\ref{localsolsec}, we shall not present them explicitly here.
  
    Some explicit examples of the Ricci-flat generalised Taub-NUT
metrics are as follows.  We set $\lambda=1$ for convenience.  For
$n=1$, $m>1$, we have
\bea
d\hat s^2_{m+4} &=&\fft{(m+1)\, dr^2}{U} +
 \fft{4U}{m+1}\, \,(d\psi+\cos\theta\, d\phi)^2 + 
(r^2-1)\, (d\theta^2+\sin^2\theta\, d\phi^2) \nn\\
&&+ (m-1)\, r^2\,
 d\Omega_m^2\,,\nn\\
U &=& 1-\fft{2 - 2 r^{1-m}}{(m-1)\,
    (r^2-1)}\,. 
\eea
The case $n=1$, $m=2$ is of some interest, giving
\be
d\hat s_6^2 = \fft{3 r\, (r+1)\, dr^2}{(r-1)\, (r+2)}
     + \fft{4(r-1)\, (r+2)}{3r\, (r+1)}\, (d\psi+\cos\theta\, d\phi)^2
+ (r^2-1)\, (d\theta^2+\sin^2\theta\, d\phi^2) + r^2\, d\Omega_2^2\,.
\label{tnut}
\ee
Comparing this with the expression for the six-dimensional generalised
Taub-BOLT metric (\ref{tbolt}), we see that the local form of one
transforms into the other under $r \rightarrow -r$.  Thus the same
local metric form (\ref{tnut}) extends smoothly onto two different
manifolds, a generalised Taub-NUT metric for $1\le r\le \infty$, and a
generalised Taub-BOLT metric for $-\infty \le r\le -2$.  This feature
was also observed for the standard six-dimensional Taub-NUT and
Taub-BOLT metrics (and for certain Spin(7) holonomy metrics) in
\cite{cglpspin7}.

     For $n=2$, $m>1$ we have
\bea
d\hat s_{m+6}^2 &=& \fft{(m+3)\, dr^2}{U} + \fft{U}{m+3}\, 
(d\tau-2A)^2 + (r^2-1)\,
d\Sigma_4^2 + (m-1)\, r^2\, d\Omega_m\,,\nn\\
U &=& 1 - 4 \fft{(m-1)\, r^2 -(m+1) + 2 r^{1-m}}{(m^2-1)(r^2-1)^2}\,,
\eea
where $d\Sigma_4^2$ is the Fubini-Study metric on $\CP^2$, 
scaled so that $\lambda=1$.

\section{Conclusions}

    In this paper, we have presented new local families of solutions
of the Einstein equations, in which the metrics
(\ref{einsttot}) and (\ref{einsttotm1}) are constructed as warped
products of $S^m$ times two-dimensional $S^2$ or $\R^2$ bundles over
an Einstein-K\"ahler space $K_{2n}$.  We have studied the circumstances under
which these metrics can extend smoothly onto compact or non-compact
manifolds.  By this means we have obtained new Einstein metrics with
positive Ricci tensor that are topologically $S^{m+2}$ bundles over
$K_{2n}$.  We have also obtained new Einstein metrics with positive
Ricci tensor on spaces that are warped products of $S^m$ times $S^2$ 
bundles over $K_{2n}$, for $m>1$.  In these latter examples, the $S^m$
itself can be replaced by any Einstein space with positive Ricci tensor.
In addition, we have obtained two new classes of complete Ricci-flat
non-compact metrics.  One class can be thought of as generalisations of
the higher-dimensional Taub-NUT metrics, on manifolds with the topologies
$S^m\times \R^{n+2}$.  The other class can be viewed as generalisations
of the higher-dimensional Taub-BOLT metrics, on manifolds with the
topologies $S^m$ times $\R^2$ bundles over $K_{2n}$.  In all these
new non-compact examples, which require $m>1$, we can replace the $S^m$
by any other Einstein space with positive Ricci tensor.

\section*{Acknowledgments}
D.N.P. thanks the George P. \& Cynthia W. Mitchell Institute for
Fundamental Physics for support and hospitality during the course of
this work.   We are very grateful to Robert Mann and Cris Stelea for
pointing out some errors in the original version of this paper.

\end{document}